\documentclass[usenatbib,useAMS]{mn2e}
\usepackage{times}
\usepackage{epsfig}

\renewcommand{\vec}[1]{\bmath{#1}}

\begin{document}
\title[(Un)predictability of microlensing events]{Parameter degeneracies and (un)predictability of gravitational microlensing events}
\author[M. Dominik]{M. Dominik,$^{1}$\thanks{Royal Society University Research Fellow}\thanks{E-mail: md35@st-andrews.ac.uk}\\
$^{1}$SUPA, University of St Andrews, School of Physics \& Astronomy, 
North Haugh, St Andrews, KY16 9SS, United Kingdom}

\maketitle

\begin{abstract}
Some of the difficulties in determining the underlying physical properties
that are relevant for observed anomalies in microlensing light curves,
such as the mass and separation of extra-solar planets orbiting the
lens star, or the relative source-lens parallax, are already anchored
in factors that limit the amount of information available from ordinary
microlensing events and in the way these are being parametrized.
Moreover, a real-time detection of deviations from an ordinary light curve while these are still in progress can only be done against a known model of the latter, and such is also required for properly prioritizing
ongoing events for monitoring in order to maximize scientific returns.
Despite the fact that ordinary microlensing light curves are described
by an analytic function that only involves a handful of parameters,
modelling these is far less trivial than one might be tempted to think.
A well-known degeneracy for small impacts, and another one for the initial rise of an event, makes an interprediction of different phases impossible,
while in order to determine a complete set of model parameters, the fundamental characteristics of all these phases need to be properly assessed. While it is found that the wing of the light curve provides valuable information
about the time-scale that absorbs the physical properties, the peak flux
of the event can be meaningfully predicted only after about a third of
the total magnification has been reached. Parametrizations based on 
observable features not only ease modelling by bringing the covariance matrix close to diagonal form, but also allow good predictions of the measured flux without the need to determine all parameters accurately.
Campaigns intending to infer planet populations from observed microlensing
events need to invest some fraction of the available time into acquiring
data that allows to properly determine the magnification function.
\end{abstract}

\begin{keywords}
gravitational lensing -- planetary systems.
\end{keywords}

\section{Introduction}

An efficient detection of planets by means of gravitational microlensing
requires sufficiently accurate predictions of the underlying ordinary
light curve against which the planetary deviations need to be identified
\citep{SIGNALMEN,ARTEMiS}. An optimal monitoring strategy in order to maximize the scientific return moreover profits strongly from the 
proper determination of a full set of model parameters as early as 
possible, and the latter becomes a requirement for finally assessing
the planet detection efficiency and drawing conclusions about the
planet population.

It is known that early-stage event prediction suffers from degeneracies
\citep{Albrow:MAP}, in particular the peak flux is hard to assess,
while the event time-scale $t_\rmn{E}$, required to
relate the observations to the underlying physical properties, can
remain strongly uncertain even after the event has been observed
over its full course, but
not covered well enough \citep{WP97}. The lack of determinacy of
$t_\rmn{E}$ is particularly apparent for strongly-blended events,
which in fact comprise the full sample for observations towards
neighbouring galaxies such as M31 \citep[e.g.][]{Baillon:M31}.

Understanding of ordinary light curves (comprising single point-like 
source and lens stars), their optimal parametrization, and the apparent degeneracies and ambiguities, is also crucial and useful 
for modelling events that involve anomalies, and drawing conclusions about e.g.\ stellar binaries, stellar masses derived from parallax measurements, and planets. Sparse event coverage in critical regions is even prone
to lead to
ordinary events allowing for multiple minima of the $\chi^2$ (least-squares) hypersurface, this fact being further complicated by the
presence of potential outliers in the data and the application of
robust-estimation techniques to deal with these \citep[e.g.][]{SIGNALMEN}.

In this paper, different phases of ordinary microlensing events along
with their characteristics are identified, and it is shown how
feature-oriented parametrizations can be used to quantify the 
behaviour while avoiding parameter correlations. Moreover, some fundamental
requirements for a monitoring strategy that allows to meet the
goal of determining either the observed flux or the corresponding
magnification are discussed. The basis for this is formed by 
approximate relations between the magnification and the angular
separation between lens and source that make the light curve independent
of the impact parameter. Together with some arising degeneracies, 
these have already been identified by
\citet{WP97}, but here some new light is shed on the implications and
the focus is on different aspects that are emergent right now, which
leads to arriving at
some new and different conclusions.

While Sect.~\ref{sec:canonical} reviews the general properties of 
ordinary microlensing events and the canonical parametrization, 
and Sect.~\ref{sec:adequate} presents the advantages of an alternative parametrization that is oriented towards the observable characteristic
features, Sect.~\ref{sec:predictability} discusses the predictability of 
ordinary microlensing events or the lack of it. A summary and final conclusions are provided in Sect.~\ref{sec:summary}.

\section{The canonical treatment of ordinary events}
\label{sec:canonical}
A microlensing event results if two stars happen to be closely aligned as seen from Earth, where 
this alignment is quantified by the unique characteristic scale of gravitational microlensing,
namely the angular Einstein radius \citep{Ein36}
\begin{equation}
\theta_\mathrm{E} = \sqrt{\frac{4GM}{c^2}\,\frac{\pi_\mathrm{LS}}{1~\mbox{AU}}}\,,
\label{eq:thetaE}
\end{equation}
where $M$ denotes the mass of the foreground lens star, $G$ is the
universal gravitational constant, $c$ is the vacuum speed of light,
and 
\begin{equation}
\pi_\mathrm{LS} = 1~\mbox{AU}\,\left(D_\mathrm{L}^{-1}-D_\mathrm{S}^{-1}
\right)
\end{equation}
stands for the relative source-lens parallax, with $D_\mathrm{L}$ 
and $D_\mathrm{S}$ being the distance from Earth to the foreground (lens) star and the observed
background source star, respectively. With $u\,\theta_\mathrm{E}$ denoting the angular separation between lens and source star, gravitational bending of light by the lens star yields an observable magnification 
of the source star as an analytic function 
of $u$, which reads
\begin{equation}
A(u) = \frac{u^2+2}{u\,\sqrt{u^2+4}}\,.
\label{eq:magnification}
\end{equation}

Stellar kinematics implies a non-vanishing relative proper motion $\vec \mu$ between lens and source star, so that $u$ is a function of time and
the magnification $A(u)$ describes a characteristic light curve.
While in the earlier history of gravitational microlensing several different event time-scales have been used \citep[e.g.][]{Griest:ML,MassMoments}, $t_\rmn{E} \equiv \theta_\rmn{E}/|\vec \mu|$ has emerged as a popular convenient choice.

\begin{figure}
\begin{center}
\includegraphics[width=84mm,clip=TRUE]{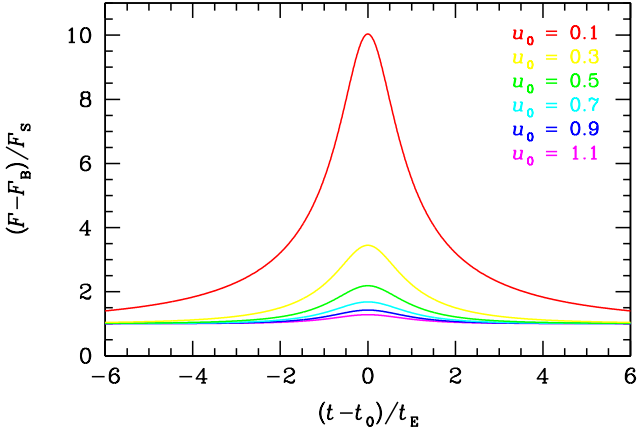}
\end{center}
\caption{Light curves for a selection of impact parameters $u_0$, 
where the event magnification $A = (F-F_\rmn{B})/F_\rmn{S}$ is plotted 
as a function of $(t-t_0)/t_\rmn{E}$, where the closest angular
approach $u_0$ is realized at $t_0$,
$t_\rmn{E} = \theta_\rmn{E}/|\vec \mu|$, with $\theta_\rmn{E}$ denoting the angular Einstein radius, defined in Eq.~(\ref{eq:thetaE}), and $\vec \mu$
being the relative proper motion between lens and source star.}
\label{fig:modellightcurves}
\end{figure}

For uniform proper motion, i.e.\ constant $\vec \mu$,
the dimensionless separation $u$ can be expressed by means of 
three quantities that
 form the parameter vector $\hat {\vec{p}} = (u_0,t_0,t_\mathrm{E})$, so that 
\begin{equation}
u (t;u_0,t_0,t_\mathrm{E}) = \sqrt{u_0^2 + \left(\frac{t-t_0}{t_\mathrm{E}}\right)^2}\,,
\label{eq:uoft}
\end{equation}
where $u_0\,\theta_\mathrm{E}$ is the smallest angular separation,
encountered at epoch $t_0$,
while the source moves by an angular Einstein radius
relative to the lens within the time-scale $t_\mathrm{E}$ \citep{Pac86}. Whereas $u_0$ and $t_0$ only define the position of the trajectory of the source relative to the lens, and therefore do not carry any information
about the relevant physical properties 
that determine the magnification,
which are the relative parallax $\pi_\mathrm{LS}$, 
the relative proper motion $\mu$,
and the lens mass $M$, all these
are convolved into
the event time-scale $t_\mathrm{E}$. It is in fact its relation to
the physical event properties that makes the parameter
$t_\mathrm{E}$ a preferred choice amongst possible time-scales. As illustrated in Fig.~\ref{fig:modellightcurves},
ordinary light curves: (1) are symmetric with respect to a peak at epoch $t_0$, (2) reach a peak flux there,
(3) approach a baseline flux for times far away from the peak,
and (4) show characteristic inflection points.
However, the parameter $t_0$ is the only one
that directly relates to the 
characteristic features of the light curve,
while all the others are not a proper reflection.
This is not at all favourable for modelling, and instead
being able to essentially read off
the model parameters from the collected data
would ease life a lot.

The phenomenon of gravitational microlensing leads to a characteristic magnification $A(t;\hat {\vec{p}})$
as a function of time $t$ and the model parameters $\hat{\vec{p}}$ that describe
the lens-observer-source geometry, the lens properties,
and the source brightness profile.
The observed flux $F^{(k)}(t; \vec{p})$ for a given site and passband
--- denoted by the multi-index $k$ ---, however, furthermore
is a linear function
of the intrinsic flux of the observed source star $F_\mathrm{S}^{(k)}$,
which is magnified,
and a background flux $F_\mathrm{B}^{(k)}$, where
\citep[e.g.][]{OB14}
\begin{equation}
F^{(k)}(t; \vec{p}) = F_\mathrm{S}^{(k)}\,A(t;\hat {\vec{p}}) + F_\mathrm{B}^{(k)}\,.
\end{equation}
This allows us to isolate these two parameters
from the remaining parameter space,
so that 
\begin{equation}
\vec{p} = (F_\mathrm{S}^{(k)}, F_\mathrm{B}^{(k)}, \hat{\vec{p}})\,,
\end{equation}
and for every value of $\hat{\vec{p}}$,
minimizing
\begin{equation}
\chi^2(t_i^{(k)}; {\vec{p}}) = \sum_{k=1}^{s} \sum_{i=1}^{n_k}
\left(\frac{F(t_i^{(k)};{\vec{p}}) - F_i^{(k)}}{\sigma_{F_i}^{(k)}}\right)^2\,,
\end{equation}
thereby obtaining a maximum-likelihood estimate of the parameter
vector $\vec p$,
means that the best-fitting source and background fluxes
can be expressed in closed analytical form \citep{RattPhd}
\begin{eqnarray}
F_\mathrm{S}  & = & \frac{\sum\frac{A(t_i)
 F_i}{\sigma_{F_i}^2} \sum\frac{1}{\sigma_{F_i}^2} -
\sum\frac{A(t_i)}{\sigma_{F_i}^2} \sum\frac{F_i}{\sigma_{F_i}^2}}
{\sum \frac{[A(t_i)]^2}{\sigma_{F_i}^2} \sum\frac{1}{\sigma_{F_i}^2}
- \left(\sum\frac{A(t_i)}{\sigma_{F_i}^2}\right)^2}\;,\nonumber \\
F_\mathrm{B} & = & \frac{\sum\frac{[A(t_i)]^2}
 {\sigma_{F_i}^2} \sum\frac{F_i}{\sigma_{F_i}^2} -
\sum\frac{A(t_i)}{\sigma_{F_i}^2} \sum\frac{A(t_i) F_i}{\sigma_{F_i}^2}}
{\sum \frac{[A(t_i)]^2}{\sigma_{F_i}^2} \sum\frac{1}{\sigma_{F_i}^2}
- \left(\sum\frac{A(t_i)}{\sigma_{F_i}^2}\right)^2}\,,
\end{eqnarray}
while the non-linear minimization process
can be restricted to
\begin{equation}
\chi^2(t_i^{(k)}, F_i^{(k)}, \sigma_{F_i}^{(k)}; \hat{\vec{p}}) = \sum_{k=1}^{s} \sum_{i=1}^{n_k}
\left(\frac{A(t_i^{(k)};\hat{\vec{p}}) - A_i^{(k)}}{\sigma_{A_i}^{(k)}}\right)^2\,,
\label{eq:chi2}
\end{equation}
expressed by means of the magnification rather than the flux, where
\begin{equation}
A_i^{(k)} = \frac{F_i^{(k)}-F_\mathrm{B}^{(k)}}{F_\mathrm{S}^{(k)}}
\end{equation}
and
\begin{equation}
\sigma_{A_i}^{(k)} = \frac{\sigma_{F_i}^{(k)}}{\left|F_\mathrm{S}^{(k)}\right|}\,.
\end{equation}

\section{Parameters that match observational features}
\label{sec:adequate}
It is well known that avoiding parameter degeneracies and strong correlations eases the modelling process. Moreover, a careful study of
these provides valuable insight into the structure of parameter space,
which can be used as a guide for developing observing strategies and
explicitly shows their limitations.
For ordinary microlensing light curves, it turns out that the observed flux can be rewritten by means of a different parametrization that better
matches the observational features, which are key to diagonalizing the
covariance matrix.

\begin{figure}
\begin{center}
\includegraphics[width=84mm,clip=TRUE]{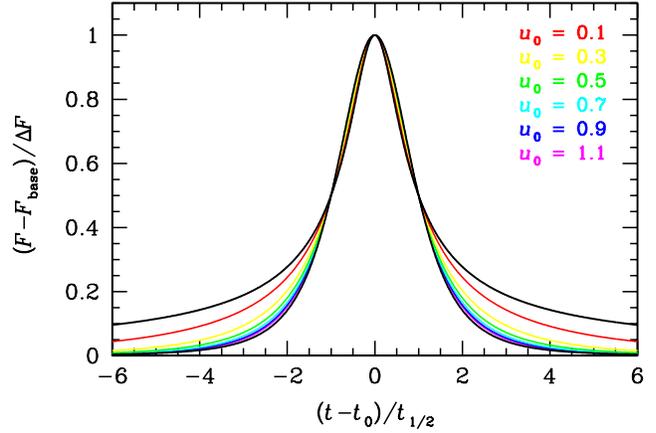}
\end{center}
\caption{Light curves for the same impact parameters as adopted
for Fig.~\protect\ref{fig:modellightcurves}, but now the relative flux
difference $(F-F_\mathrm{base})/\Delta F$ with respect to its peak
is plotted as a function of $(t-t_0)/t_{1/2}$, where the half-maximum
time-scale $t_{1/2}$ is defined by Eq.~(\protect\ref{eq:defthalf}).
All light curves range between the asymptotics for $u \ll 1$ and $u \gg 1$
(shown in black),
as given by Eq.~(\protect\ref{eq:asymptotics}).}
\label{fig:lightcurves2}
\end{figure}

With the baseline flux $F_\mathrm{base}^{(k)}$
and the peak flux $F_0^{(k)}$ given by
\begin{eqnarray}
F_\mathrm{base}^{(k)} & = & F_\mathrm{S}^{(k)}+F_\mathrm{B}^{(k)}
\,, \nonumber \\
F_0^{(k)} & = & F_\mathrm{S}^{(k)}\,A[u(t_0;\hat{\vec{p}})] +F_\mathrm{B}^{(k)}\,,
\end{eqnarray}
the maximal flux difference $\Delta F^{(k)}$ reads 
\begin{equation}
\Delta F^{(k)} = F_0^{(k)}-F_\mathrm{base}^{(k)} = F_\mathrm{S}^{(k)}\,[A(u_0)-1]\,.
\end{equation}
Not only in $F_\mathrm{S}^{(k)}$ and $F_\mathrm{B}^{(k)}$, but also
in $F_\mathrm{base}^{(k)}$ and $\Delta F^{(k)}$, the observed flux
$F^{(k)}(t; \vec{p})$ is a linear function, namely
\begin{equation}
F^{(k)}(t; \vec{p}) = \Delta F^{(k)}\,\frac{A[u(t;\hat{\vec{p}})]-1}{A(u_0)-1}
\;+\; F_\mathrm{base}^{(k)}\,,
\end{equation}
so that these parameters can be separated by a linear fit as well.
One can then define a half-maximum time $t_{1/2}$,
so that at time $t_0 \pm t_{1/2}$, 
half of the flux offset is encountered, i.e.\ 
\begin{equation}
F^{(k)}(t_0 \,\pm\, t_{1/2};u_0, t_0, t_\mathrm{E})
-F_\mathrm{base}^{(k)} = \frac{1}{2}\,\Delta F^{(k)}\,.
\label{eq:defthalf}
\end{equation}
This means that the epochs $t_0 \,\pm\, t_{1/2}$ correspond to a magnification
\begin{equation}
A_{1/2}(u_0) \equiv A[u(t_0 \,\pm\, t_{1/2};u_0, t_0, t_\mathrm{E})] = \frac{A(u_0)+1}{2}\,.
\end{equation}
With $u_{1/2} \equiv u(t_0 \,\pm\, t_{1/2};u_0, t_0, t_\mathrm{E})$
and by means of Eq.~(\ref{eq:uoft}), one finds
\begin{equation}
t_{1/2} = t_\mathrm{E}\,
\sqrt{u_{1/2}^2-u_0^2}\,,
\end{equation}
which leads to
\begin{equation}
u(t; u_0, t_0, t_{1/2}) = u_0\,\sqrt{1+ \left(\frac{u_{1/2}^2}{u_0^2} - 1
\right)\,\left(\frac{t-t_0}{t_{1/2}}\right)^2}\,.
\end{equation}
If one now plots the offset flux $F^{(k)}(t; \vec{p}) -F_\mathrm{base}^{(k)}$
in units of the difference $\Delta F^{(k)}$ between peak and baseline,
and scales $t-t_0$ with $t_{1/2}$ rather than $t_\mathrm{E}$,
as illustrated in Fig.~\ref{fig:lightcurves2},
one finds that light curves with different $u_0$
nearly coincide around their peaks. In contrast, it is the wing region
of the event $1.5~t_{1/2} \la |t-t_0| \la 3~t_{1/2}$ that is best-suited
to provide information about the impact parameter $u_0$, and thereby
of the event time-scale $t_\rmn{E}$, related to the underlying physical
properties.\footnote{The subsequent section shows that the asymptotic degeneracy for large $u$ does not extend into this region.}
In principle, $u_0$ is determined by the slope of the light curve at
$t = t_0 \pm t_{1/2}$, but the narrow range\footnote{The absolute value of the slope ranges between 0.375 (for $u_0 \to 0$) and $2-\sqrt{2} \approx 0.586$ (for $u_0 \to \infty$).} strongly limits
the feasibility of such an approach in practice.

\begin{table*}
\caption{Asymptotic behaviour of the observed flux $F^{(k)}$, as well
as of other relevant quantities, in the limits 
$u \ll 1$ 
or $u \gg 1$, and its matching parametrization.}
\label{tab:asymptotics}
\small
\begin{minipage}{165cm}
\begin{tabular}{@{}lcccc}
\hline\\[-0.35pc]
 &  & $u \ll 1$ & $\qquad$ & $u \gg 1$\\[0.3pc] \hline\\[-0.3pc]
$A(u)$ & $\simeq$\quad & $\displaystyle u^{-1}$ & $\qquad$ & $\displaystyle 1 + 2\,u^{-4}$\\[0.35pc]
$\displaystyle \frac{A[u(t;\hat {\vec{p}})] - 1}{A(u_0)-1}$ & $\simeq$ &
$\displaystyle \frac{u_0}{u(t)}$ & \qquad &
$\displaystyle 
\left(\frac{u_0}{u(t)}\right)^4$\\[1.05pc]
$A_{1/2}$ & $\simeq$ & $\displaystyle \frac{1}{2\,u_0}$ & $\qquad$ & 
$\displaystyle 1 + \frac{1}{u_0^4}$\\[1.05pc]
$u_{1/2}$ & $\simeq$ & $2\,u_0$ & \qquad & $\sqrt[4]{2}\,u_0$\\[0.35pc]
$t_{1/2}$ & $\simeq$ &  $\sqrt{3}\,u_0\,t_\mathrm{E}$ & \qquad &
$(\sqrt{2}-1)\,u_0\,t_\mathrm{E}$\\[0.5pc]
$u(t)$ & $\simeq$ & $u_0\,\sqrt{1+3\,\left(\displaystyle\frac{t-t_0}{t_{1/2}}\right)^2}$
& \qquad &  $u_0\,\sqrt{1+(\sqrt{2}-1)\,\left(\displaystyle\frac{t-t_0}{t_{1/2}}\right)^2}$ 
\\[1.35pc]
$\displaystyle F^{(k)}(t; 
\Delta F^{(k)},  F_\mathrm{base}^{(k)},
t_0,t_{1/2})-F_\mathrm{base}^{(k)}$ & $\simeq$ & $\displaystyle \frac{\Delta F^{(k)}}
{\sqrt{1+3\,\left(\displaystyle\frac{t-t_0}{t_{1/2}}\right)^2}}$ & \qquad &
$\displaystyle \frac{\Delta F^{(k)}}
{\left[1+(\sqrt{2}-1)\,\left(\displaystyle\frac{t-t_0}{t_{1/2}}\right)^2\right]^2}$ \\
\\[-0.3pc]
\hline
\end{tabular}
\end{minipage}
\end{table*}

It is also instructive to explore the limits of small and
large separations more closely. Both for $u \ll 1$ and $u \gg 1$,
the observed relative flux offset $(F^{(k)}(t)-F_\rmn{base}^{(k)})/\Delta F^{(k)}$ becomes independent of $u_0$, while $t_{1/2}$ becomes
proportional to $u_0\,t_\mathrm{E}$, however with different proportionality
factors for the two extreme cases. As summarized in
Table~\ref{tab:asymptotics}, this is due to the fact that the magnification can reasonably well be approximated by a more simple expression, namely $[A(u)-1]/[A(u_0)-1]$ approaching
$u_0/u$ for $u \ll 1$, and $(u_0/u)^4$ for $u \gg 1$.
For $u \ll 1$, one retrieves the known results that apply
to microlensing of unresolved sources where $F_\rmn{B}^{(k)} \gg F_\rmn{S}^{(k)}$, being an inevitability for observations towards
M31 or other nearby other galaxies \citep{Baillon:M31,WP97}.
Looking at Fig.~\ref{fig:lightcurves2} again, one sees that all light curves range between the two extreme cases
\begin{equation}
\frac{F^{(k)}(t)-F_\rmn{base}^{(k)}}{\Delta F^{(k)}} \simeq
\left\{ \begin{array}{l}
%\displaystyle
\left[1+(\sqrt{2}-1)\,\left(\frac{t-t_0}{t_{1/2}}\right)^2\right]^{-2}
\;  (u \gg 1) \,,  \\[3.5ex]
\vspace*{-0.2ex}
%\displaystyle 
\left[1+3\,\left(\frac{t-t_0}{t_{1/2}}\right)^2\right]^{-1/2}
\qquad (u \ll 1)\,.
\end{array}\right.
\label{eq:asymptotics}
\end{equation}
The wings of the light curves converge towards the expression for $u \gg 1$, where for smaller $u_0$, the transition from the $u \ll 1$ to
the $u \gg 1$ asymptotic occurs at larger $|t-t_0|/t_{1/2}$.

\section{Different event phases and the lack of predictability}
\label{sec:predictability}
One might think that 3 parameters (like $u_0$, $t_0$, and $t_\mathrm{E}$) can be obtained straightforwardly by means of regression from as few as 4 data points, but such an attempt fails if the observable does not significantly change with a variation of the considered parameter. So far, it has been assumed that the light curve has been sampled over its full course, so that a full set of characteristics can be determined from which model parameters
can be derived. However, the light curve develops in time, so that some characteristics are not accessible at
early stages. While the baseline flux $F_\rmn{base}^{(k)}$ is being observed much before any rise in brightness occurs,
the peak flux $F_0^{(k)}$ or the flux shift $\Delta F^{(k)} = F_0^{(k)}-F_\rmn{base}^{(k)}$, respectively, remain unknown, as we shall
see more explicitly in the following.

Reviewing the asymptotics for $u \ll 1$ and $u \gg 1$ from a different perspective already shows us that there is no interpredictability 
between the peak and wing regions of the observed light curve,
and in particular,
early observations give poor estimates of the peak magnification,
as well as on the time-scale $t_\mathrm{E}$. In particular, maximum-likehihood estimates corresponding to values that minimize $\chi^2$,
Eq.~(\ref{eq:chi2}), frequently yield very small $u_0$, far
away from expectations. Therefore, \citet{Albrow:MAP} has suggested
to use a maximum-a-posteriori estimate instead, incorporating the actual distribution of the parameters of the observed events of the microlensing surveys as prior.

Let us look into this with a view on the parameter degeneracies.
The earliest stages of an event are characterized by $|t-t_0|/t_\rmn{E} \gg u_0$, so that
$u \simeq |t-t_0|/t_\rmn{E}$. For the two extreme cases of $u$ being far from unity, one finds
\begin{equation}
A(u) \simeq \left\{ \begin{array}{l} 
\displaystyle
1 + 2\,\left(\frac{t_\rmn{E}}{|t-t_0|}\right)^4 \quad (u \gg u_0, u \gg 1)\,,  \\ \\
\displaystyle
\frac{t_\rmn{E}}{|t-t_0|} \qquad (u_0 \ll u \ll 1)\,,
\end{array}\right.
\end{equation}
where the second case is realized for a substantial time interval
if $u_0$ is sufficiently small.

With the blend ratio $g^{(k)} = F_\rmn{B}^{(k)}/F_\rmn{S}^{(k)}$,
the observed flux $F^{(k)}(t)$ can be written as
\begin{eqnarray}
F^{(k)}(t) & = & F_\rmn{base}^{(k)}\,\left[\frac{A[u(t;\hat{\vec p})]+g^{(k)}}{1+g^{(k)}}\right] \nonumber \\
  & = & F_\rmn{base}^{(k)}\,\left[\frac{A[u(t;\hat{\vec p})]-1}{1+g^{(k)}}+1\right]\,,
\end{eqnarray}
so that
\begin{equation}
\frac{F^{(k)}(t)}{F_\rmn{base}^{(k)}} -1 \simeq
\left\{ \begin{array}{l}
\displaystyle
\frac{2}{1+g^{(k)}}\,\left(\frac{t_\rmn{E}}{|t-t_0|}\right)^4 = 2\,\left(\frac{t_\rmn{r}^{(k)}}{|t-t_0|}\right)^4\\[2.5ex]
\qquad (u \gg u_0, u \gg 1) \,,\\[3.5ex]
\vspace*{-0.2ex}
\displaystyle
\frac{1}{1+g^{(k)}}\,\frac{t_\rmn{E}}{|t-t_0|} = \frac{t_\rmn{s}^{(k)}}{|t-t_0|}\\[2.5ex]
\qquad (u_0 \ll u \ll 1)\,,
\end{array}\right.
\label{eq:approx}
\end{equation}
where 
\begin{equation}
t_\rmn{r}^{(k)} \equiv \frac{t_\rmn{E}}{\sqrt[4]{1+g^{(k)}}}\,,
\qquad
t_\rmn{s}^{(k)} \equiv \frac{t_\rmn{E}}{1+g^{(k)}}\,,
\label{eq:risedef}
\end{equation}
are characteristic rise times, absorbing $t_\rmn{E}$ and $g^{(k)}$,
with $t_\rmn{r} = t_\rmn{s} = t_\rmn{E}$ for $g=0$.

\begin{figure}
\begin{center}
\includegraphics[width=84mm,clip=TRUE]{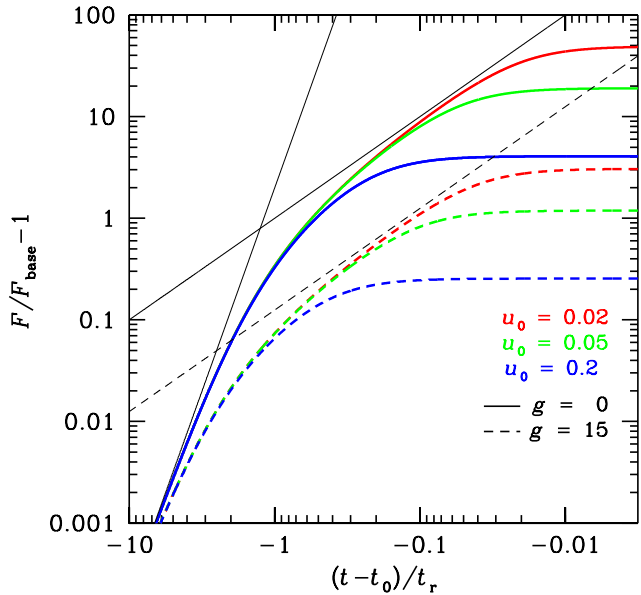}
\end{center}
\caption{Development of a microlensing event from a common initial rise to
different peak fluxes for three selected impact parameters $u_0$ (colour-coded) and two blend ratios $g^{(k)} \equiv F_\rmn{B}^{(k)}/F_\rmn{S}^{(k)}$ (solid or dashed line). Double-logarithmic plot showing the relative brightening
$F^{(k)}(t)/F_\rmn{base}^{(k)}-1$ above baseline as a function
of $(t-t_0)/t_\rmn{r}^{(k)}$, where $t_\rmn{r}^{(k)}$ is the rise time
defined by Eq.~(\protect\ref{eq:risedef}), so that for the two
selected blend ratios $t_\rmn{E} = t_\rmn{r}^{(k)}$ ($g^{(k)} = 0$)
or $t_\rmn{E} = 2~t_\rmn{r}^{(k)}$ ($g^{(k)} = 15$), respectively. The black lines correspond to
the asymptotic behaviour given by Eq.~(\ref{eq:approx}).
Be aware of the fact that the same scale corresponds to tiny changes
in the observed flux near the bottom of the plot, but huge ones near the top. Similarly, the left parts span large time intervals, while the right
parts span small ones.}
\label{fig:earlycurves}
\end{figure}

Figure~\ref{fig:earlycurves} shows $F^{(k)}(t)/F_\rmn{base}^{(k)}-1$
as a function of $(t-t_0)/t_\rmn{r}$ in a double-logarithmic plot,
so that the approximate relations of Eq.~(\ref{eq:approx}) correspond
to straight lines. It illustrates that different behaviour
allows to distinguish three phases of a microlensing event, corresponding
to the initial rise, a mid-phase, and the peak approach, where a
determination of the full model parameter set requires an assessment
of the fundamental characteristics of all these phases.
With the approximate $F^{(k)}(t)$ diverging as $t \to t_0$, an estimate for $t_0$ based on such an approximation corresponds
to $u_0 = 0$. Only a departure from the approximation in the form of
an evident turn-off gives evidence for non-vanishing $u_0$. However,
 as Fig.~\ref{fig:earlycurves} illustrates, this happens at rather
late stage, and roughly when a third of the peak magnification has been reached. Given that for $u \la 1$, $A(u)$ differs substantially from the 
asymptotic behaviour for $u \gg 1$, the blend ratio $g^{(k)}$ can in
principle be determined rather early, which then
provides the time-scale $t_\rmn{E}$ that is related to the underlying physical properties of the event. In practice this
requires sufficiently dense and precise measurements, which are frequently
not available for fluxes close to $F_\rmn{base}$, but the long duration
of the rise phase in principle allows for lots of data to be collected.
 Nevertheless, as soon
as both $t_\rmn{r}^{(k)}$ and $t_\rmn{s}^{(k)}$ can be determined from the
acquired data, the blend ratio $g^{(k)}$ and $t_\rmn{E}$ are known.
Again, one sees the power of observations covering the wing of the light curve.

\begin{figure}
\begin{center}
\includegraphics[width=84mm,clip=TRUE]{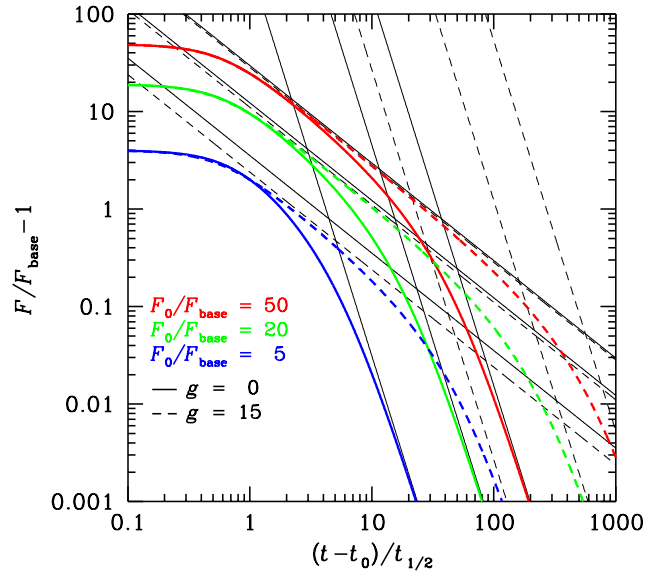}
\end{center}
\caption{
Development of events from the peak with a given relative brightening 
$F_0^{(k)}/F_\rmn{base}^{(k)}$, where light curves corresponding to
three selected values are shown, towards the baseline flux
$F_\rmn{base}^{(k)}$ for two different blend ratios $g \equiv F_\rmn{S}^{(k)}/F_\rmn{B}^{(k)}$. Given that the peak region is characterized 
by its width, $F_0^{(k)}/F_\rmn{base}^{(k)}-1$ is shown as function
of $(t-t_0)/t_{1/2}$ (in a double-logarithmic plot). As for Figure~\ref{fig:earlycurves}, the asymptotic
behaviour given by Eq.~(\ref{eq:approx}) is shown by means of
black lines.}
\label{fig:halfcurves}
\end{figure}

Going the other way round and coming from the peak, the 
microlensing light curve first follows the
decay with $t_{1/2}$, then enters the mid-phase characterized by $t_\rmn{s}^{(k)}$, and finally follows a decrease described by $t_\rmn{r}^{(k)}$. Double-logarithmic plots of the relative offset 
brightening  $F_0^{(k)}/F_\rmn{base}^{(k)}-1$ as a function of 
$(t-t_0)/t_{1/2}$ are shown in Fig.~\ref{fig:halfcurves}, while the
relevant magnifications, impact parameters, and time-scales for the selected cases are listed in Table~\ref{tab:prop}. Again, one sees that
the light curves for different blend ratio and $t_\rmn{E}$ can be made
to match well near the peak, whereas observations in the wing of the light
curve on the departure from the asymptotic mid-phase behaviour ($u_0 \ll u \ll 1$) or on the transition into the baseline-approach phase ($u \gg 1$)
are useful to determine $t_\rmn{E}$. The rather long duration of these
phases favours such an attempts by not requiring a high sampling rate 
for obtaining a larger number of measurements.

\begin{table*}
\caption{Magnifications, impact parameters and time-scales for the
relative peak fluxes adopted in Fig.~\protect\ref{fig:halfcurves}.}
\label{tab:prop}
\begin{minipage}{180mm}
\begin{tabular}{@{}cccccccc}
\hline
$F_0^{(k)}/F_\rmn{base}^{(k)}$ & $A_0$ &  $A_{1/2}$ & 
$u_0$ & $u_{1/2}$  &  $t_\rmn{E}/t_{1/2}$ &  $t_\rmn{r}^{(k)}/t_{1/2}$  &  $t_\rmn{s}^{(k)}/t_{1/2}$ \\ \hline
\multicolumn{8}{c}{$g=0$} \\
\hline
50 & 50 & 25.5 & 0.0200 & 0.0392 & 29.6 & 29.6 & 29.6 \\
20 & 20 & 10.5 & 0.0500 & 0.0956 & 12.3 & 12.3 & 12.3 \\
5 & 5 & 3 & 0.203 & 0.348 & 3.53 & 3.53 & 3.53 \\ \hline
\multicolumn{8}{c}{$g=15$} \\ \hline
50 & 785 & 393 & 0.00127 & 0.00255 & 453 & 227 & 28.4 \\
20 & 305 & 153 & 0.00328 & 0.00654 & 177 & 88.4 & 11.1 \\
5 & 65 & 33 & 0.0154 & 0.0303 & 38.3 & 19.1 & 2.39 \\
\hline
\end{tabular}

\medskip
\footnotesize
$t_\rmn{E} = 2~t_\rmn{r}^{(k)} = 16~t_\rmn{s}^{(k)}$ for $g = 15$, while
$t_\rmn{E} = t_\rmn{r}^{(k)} = t_\rmn{s}^{(k)}$ for $g = 0$.
\end{minipage}
\end{table*}

\section{Summary and final conclusions}
\label{sec:summary}
Despite the fact that the timescale $t_\rmn{E} \equiv \theta_\rmn{E}/\mu$,
where $\theta_\rmn{E}$ denotes the angular Einstein radius, and $\mu$
the relative proper motion between lens and source star, is the crucial one for drawing conclusions about the underlying physical
properties that led to a microlensing event, it is not a good choice for
describing the observable characteristic features, and thereby not 
a useful means of describing or predicting events in progress.

Contrary to common belief, 3 model parameters cannot always properly
be extracted from a least-squares fit involving at least 4 data points.
In fact, such attempts fail if the respective function to be matched
to the observed data does not significantly depend on each of the
parameters over the region where data have been acquired.
Microlensing light curves usually go through 3 phases from baseline
to peak as well as from peak back to baseline: Two rise phases
characterized by different rise times $t_\rmn{r}^{(k)}$ and 
$t_\rmn{s}^{(k)}$, corresponding to different power laws of the
magnification with the angular separation between lens and source,
as well as a peak region, characterized by a half-width $t_{1/2}$.
None of the individual phases contains characteristic information about the
complete set of model parameters, and in order to reveal accurate estimates
for all of them, each of them requires appropriate coverage.
While the mid-phase gets squeezed for impact angles $u\,\theta_\rmn{E}$
of the order of the angular Einstein radius $\theta_\rmn{E}$ or larger,
the part of the light curve with $u \gg 1$ becomes indistinguishable from the baseline for strongly-blended events, so
that on the approach to baseline, the offset magnification is 
proportional to $u^{-1}$ rather than $u^{-4}$, given that still
$u \ll 1$ for $|t-t_0| \gg 1$.

It is well-known that an appropriate estimate of event parameters at
early event stages is not feasible, and in particular the peak magnification is regularly overpredicted by a maximum-likelihood estimate
corresponding to minimizing the sum of normalized squared deviations $\chi^2$. Just for this reason, \citet{Albrow:MAP} had suggested to 
use a maximum-a-posteriori estimate instead, with a suitable prior.
While this brings the estimate closer to its expectation value, it
does not get around the uncertainty. A closer eximation shows that
the light curve is compatible with an infinite peak flux until roughly
a third of the true offset magnification is reached. Moreover,
it is a wing region $1.5~t_{1/2} \la |t-t_0| \la 3~t_{1/2}$ that is
best suited to determine the blend ratio $g^{(k)} = F_\rmn{B}^{(k)}/F_\rmn{S}^{(k)}$ (and with it the time-scale $t_\rmn{E}$, rather than 
the immediate vicinity of the peak. The rather long duration of this 
phase allows to obtain a suitable measurement without the need for
very dense sampling.

For an accurate prediction of the observed flux, a proper determination
of the full set of model parameters is not required, so that local
approximations can provide a reasonable substitute.
In sharp contrast,
proper knowledge of $t_\rmn{E}$, which implies knowledge of 
the blend ratio $g^{(k)}$ and the magnification $A(t)$, is a 
requirement for determining the event detection efficiency to
planets \citep{fivelong} as well as for prioritising ongoing events in
order to maximize it \citep{Han:priorities,webPLOP}. Without such
information, one neither knows the amplitude, nor the duration, nor
the location of potentially arising planetary signals.
Therefore, an efficient campaign for inferring the planet population
from observed microlensing events needs to invest time into observations
that allow to properly determine the event parameters, rather than just
trying to detect planets in poorly determined events, where unsuitable
assumptions about model parameters may yield to bad choices, or efforts
could even turn out to be wasted if the planet detection efficiency 
cannot be assessed. Building upon the findings presented in this paper, 
a more detailed study of event (un)predictability taking into account
the specific capabilities of observing campaigns could hence provide
important clues towards optimizing strategies for detecting planets
and determining their population statistics.

\bibliographystyle{mn2e}
\bibliography{unpredict}

\end{document}